\documentclass[a4paper,10pt]{article}
\usepackage{amsfonts}
\usepackage{amsmath}
\usepackage{amssymb}
\usepackage{graphicx}
\usepackage{caption}

\begin{document}

\title{Magnetization in pristine graphene with Zeeman splitting and variable
spin-orbit coupling}
\author{F. Escudero$^{\dag }$, L. Sourrouille$^{\dag }$, J.S. Ardenghi$^{\dag }$%
\thanks{%
email:\ jsardenghi@gmail.com, fax number:\ +54-291-4595142} and P. Jasen$%
^{\dag }$ \\
$^{\dag }$IFISUR, Departamento de F\'{\i}sica (UNS-CONICET)\\
Avenida Alem 1253, Bah\'{\i}a Blanca, Buenos Aires, Argentina}
\maketitle

\begin{abstract}
The aim of this work is to describe the spin magnetization of graphene with
Rashba spin-orbit coupling and Zeeman effect. It is shown that the
magnetization depends critically on the spin-orbit coupling $\lambda $ that
is controlled with an external electric field. In turn, by manipulating the
density of charge carriers, it is shown that spin up and down Landau levels
mix introducing jumps in the spin magnetization. Two magnetic oscillations
phases are described that can be tunable through the applied external
fields. The maximum and minimum of the oscillations can be alternated by
taking into account how the energy levels are filled when the
Rashba-spin-orbit coupling is turned on. The results obtained are of
importance to design superlattices with variable spin-orbit coupling with
different configurations in which spin oscillations and spin filters can be
developed.
\end{abstract}

\section{ \ Introduction}

Graphene, a one-atom-thick allotrope, has become one of the most significant
topics in solid state physics due to its two-dimensional structure as well
as from its unique electronic properties (\cite{novo},\cite{intro1},\cite%
{intro2}, \cite{B}, \cite{BBBB}). The carbon atoms form a honey-comb lattice
made of two interpenetrating triangular sublattices, $A$ and $B$ that create
specific electronic band structure at the Fermi level: electrons move with a
constant velocity about $c/300$. In turn, the electronic properties are
dictated by the $\pi $ and $\pi ^{\prime }$ bands that form conical valleys
touching at the two independent high symmetry points at the corner of the
Brillouin zone, the so called valley pseudospin \cite{A}. In the absence of
defects, electrons near these symmetry points behave as massless
relativistic Dirac fermions with an effective Dirac-Weyl Hamiltonian \cite{B}
which allows to consider graphene as a solid-state toy for relativistic
quantum mechanics. When the interaction between the orbital electron motion
and spin degrees of freedom is taken into account, the spin-orbit coupling
(SOC) induces a gap in the spectrum. SOC is the most important interaction
affecting electronic spin transport in nonmagnetic materials. The use of
graphene in spintronics (\cite{zutic} and \cite{fabian}) would require
detailed knowledge of graphene's spin-orbit coupling effects, as well as
discovering ways of increasing and controlling them. The SOC in graphene
consists of intrinsic and extrinsic components (\cite{kons} and \cite%
{balakri}). The intrinsic component of SOC is weak in the graphene sheet,
although considerable magnitudes can be obtained in nanoscaled graphene (%
\cite{kane1} and \cite{kane2}). The extrinsic component, known as the Rashba
spin-orbit coupling (RSOC) \cite{bich} can be larger than the intrinsic
component \cite{deck}. The RSOC may be controlled by the application of
external electric fields (\cite{ert} and \cite{aba}). On the other side,
when a magnetic field is applied perpendicular to the graphene sheet, a
discretization of the energy levels is obtained, the so called Landau levels 
\cite{kuru}. These quantized energy levels still appear also for
relativistic electrons, just their dependence on field and quantization
parameter is different. In a conventional non-relativistic electron gas,
Landau quantization produces equidistant energy levels, which is due to the
parabolic dispersion law of free electrons. In graphene, the electrons have
relativistic dispersion law, which strongly modifies the Landau quantization
of the energy and the position of the levels. In particular, these levels
are not equidistant as occurs in a conventional non-relativistic electron
gas in a magnetic field. This large gap allows one to observe the quantum
Hall effect in graphene, even at room temperature \cite{E}. In turn, when
magnetic fields are applied to solids an important effect called the de
Haas-van Alphen (dHvA) \cite{haas} appears as oscillations of magnetization
as a function of inverse magnetic field. This effect is a purely quantum
mechanical phenomenon and is a powerful toof for maping the Fermi surface,
i.e. the electronic states at the Fermi energy (\cite{meinel},\cite{schwarz}%
) and because gives important information on the energy spectrum, it is one
of the most important tasks in condensed matter physics. The different
frequencies involved in the oscillations are related to the closed orbits
that electrons perform on the Fermi surface. It has been predicted in
graphene that magnetization oscillates periodically in a sawtooth pattern,
in agreement with the old Peierls prediction \cite{shara}, although the
basic aspects of the behavior of the magnetic oscillations for quasi-2D
materials remains yet unclear \cite{champel}. In contrast to 2D conventional
semiconductors, where the oscillating center of the magnetization $M$
remains exactly at zero, in graphene the oscillating center has a positive
value because the diamagnetic contribution is half reduced with that in the
conventional semiconductor \cite{fzhen}. From an experimental point of view,
carbon-based materials are more promising because the available samples
already allows one to observe the Shubnikov-de Haas effect (\cite{uji} and 
\cite{wang}) and then may be easier to interpret than quantum oscillations
in its transport properties. Because the dHvA signal in 2D systems are free
of the $k_{z}$ smearing, it should be easier to obtain much rich information
about the electron processes. In addition, the SOC, which is considerably
large in graphene \cite{vary} plays an important role in the determination
of the magnetic oscillations because of the fundamental difference with
conventional semiconductor 2DEG.

Motivated by this phenonema, in the present paper we study the dHvA
oscillations in the spin magnetization in graphene by taking into account
the Rashba spin-orbit interaction modulated by a perpendicular external
electric field and the Zeeman effect modulated by a constant magnetic field
perpendicular to the graphene sheet. It is well known that each electron
contributes with $\mu _{B}$ to the magnetization density if the spin is
parallel to the applied magnetic field and $-\mu _{B}$ if it is
antiparallel. Hence, if $N_{\pm }$ is the number of electrons per area with
spin parallel ($+$) or antiparallel ($-$), the magnetization density will be 
$M=\mu _{B}(N_{+}-N_{-})$. As it was said before, the introduction of a
constant magnetic field introduces the Landau levels that are splitted by
the Zeeman effect. The filling of these levels is not trivial in graphene
due to its square root dependence in the Landau index. The degeneracy, that
depends linearly on $B$, defines the number of states that are completely
filled, but these levels may not be sorted in ascending order with
intercalated spins. In fact, whenever the $n+1$ Landau level with spin up is
lower than the $n$ Landau level with spin down, an enhancement of the spin
magnetization will be obtained. This behavior can be alter drastically with
the RSOC, because in this case, the ordering of the energy levels is not
trivial. In this sense and since both the magnetic field and RSOC\ are
externally controllable, in what follows the role of each of the parameters
involved, as the electron density and the electric and magnetic field
strength are discussed in relation to the maximum and minimum of the
oscillations. The spin-orbit effects are important, besides the fundamental
electronic and band structure and its topology, to understand spin
relaxation, spin Hall effect and other effects such as weak
(antilocalization). There are two main routes to implement spintronics
devices: the giant magneto resistance effect (GMR) \cite{Grun} and the spin
effect field transistor (spin-FET)\ \cite{datta}. Both devices consists in a
sandwich structure made of two ferromagnetic materials separated by a
non-magnetic later in GMR and two dimensional electron gas in spin-FET. In
this device, the spin-orbit coupling causes the electron spin to precess
with a precession length determined by the strength of the spin-orbit
coupling, which through a gate voltage becomes tunable. A detailed knowledge
of the interplay of the the different parameters entering in the Hamiltonian
is vital to fundamentally understand the spin-dependent phenomena in
graphene.\footnote{%
For a good review of spintronics in graphene see \cite{kheira}.} This work
will be organized as follow: In section II, pristine graphene under a
constant magnetic field with spin-orbit coupling and Zeeman splitting is
introduced and the magnetic oscillations are discussed. In section III, $%
N_{+}$ and $N_{-}$ populations as a function of $B$, the electron density
and RSOC\ parameter is studied and the oscillations are computed and
discussed. The principal findings of this paper are highlighted in the
conclusion.

\section{Magnetic oscillations with Zeeman splitting}

For a self-contained lecture of this paper, a brief introduction of the
quantum mechanics of graphene in a constant magnetic field in the long
wavelength approximation will be introduced (see \cite{BBBB}). The
Hamiltonian in one of the two inequivalent corners of the Brillouin zones
can be put in a compact notation as (see \cite{rash} and \cite{fzhen}) 
\begin{equation}
H=v_{F}(\mathbf{\sigma \cdot \pi \mathbf{)}}+\frac{1}{2}\Delta _{R}(\mathbf{%
\sigma \times \mathbf{s)}}-\Delta _{Z}s_{z}  \label{a1}
\end{equation}%
where $\mathbf{\sigma }$ are the Pauli matrices acting on the pseudospin
space and $\mathbf{s}$ are the Pauli matrices acting on the spin space. $%
\Delta _{R}$ is the extrinsic spin-orbit coupling that arise when an
external electric field is applied perpendicular to the graphene sheet or
from a gate voltage or charged impurities in the substrate (\cite{wu}, \cite%
{manja} and \cite{bur}).\footnote{%
Intrinsic spin-orbit coupling $\Delta _{int}$ that contains contributions
from the $p$ orbitals and $d$ orbitals (see eq.(10) of \cite{kons})is
neglected.} $\Delta _{Z}=\frac{g\mu _{B}}{2}B$ is the Zeeman energy that
depends on the magnetic field strength. In the following we will write the
extrinsic Rashba spin orbit coupling as $\Delta _{R}=\hbar \omega _{y}$. The
quasiparticle momentum must be replaced by $\mathbf{\pi =p-}e\mathbf{A}$,
where $e$ is the electron charge and $\mathbf{A}$ is the vector potential
which in the Landau gauge reads $\mathbf{A=}(-By,0,0)$. For the $K$ valley,
the third term can be written as $-i\Delta _{R}(\sigma _{+}s_{+}-\sigma
_{-}s_{-})$ where $\sigma _{\pm }=\sigma _{x}\pm i\sigma _{y}$ and $s_{\pm
}=s_{x}\pm is_{y}$. This term describes a coupling between pseudospin and
spin states, that is, a spin-flip process can be achieved by hopping an
electron in the $A$ sublattice with spin up to the $B$ sublattice with spin
down. The eigenvalues of the Hamiltonian of eq.(\ref{a1}) has been computed
in \cite{rash} and \cite{fzhen} without the Zeeman term. By introducing this
term the eigenvalues reads\footnote{%
In \cite{tsaran} it is discussed the intrinsic and extrinsic spin-orbit
couplings and the quasi-classical limit for higher Landau levels.} 
\begin{equation}
E_{n,s,l}=\frac{l\hbar }{\sqrt{2}}\sqrt{2\omega _{Z}^{2}+2n\omega
_{L}^{2}+\omega _{y}^{2}-s\sqrt{16n\omega _{Z}^{2}\omega _{L}^{2}+4n\omega
_{L}^{2}\omega _{y}^{2}+\omega _{y}^{4}}}  \label{aa1.3}
\end{equation}%
where $l=+1(-1)$ is for the conduction (valence) band, $n=0,1,2,...$ is the
Landau level index and $s=\pm 1$ is the spin. The number of conduction
electrons will be given by $N=n_{e}A$ where $n_{e}$ is the electron density
and $A$ is the area of the graphene sheet. For simplicity, we can consider
only conduction electrons, which implies that the added electrons are due to
a gate voltage applied to the graphene sheet so that $n_{e}$ can be varied
as a function of $V_{G}$. When a magnetic field is applied, the energy is
discretized and each level is degenerated with degeneracy $D=BA/\phi $,
where $\phi $ is the magnetic unit flux. There is a critical magnetic field
in which the degeneracy $D$ equals the number of electrons $N$%
\begin{equation}
B_{C}=n_{e}\phi  \label{n1}
\end{equation}%
If we would consider only valence electrons, then $B_{C}\sim 80\times
10^{3}T $, which is an unfeasible experimentally. Nevertheless, $B_{C}$
defined as in last equation will serve as an upper bound for the numerical
calculations.

\begin{figure}[tbp]
\centering\includegraphics[width=120mm,height=70mm]{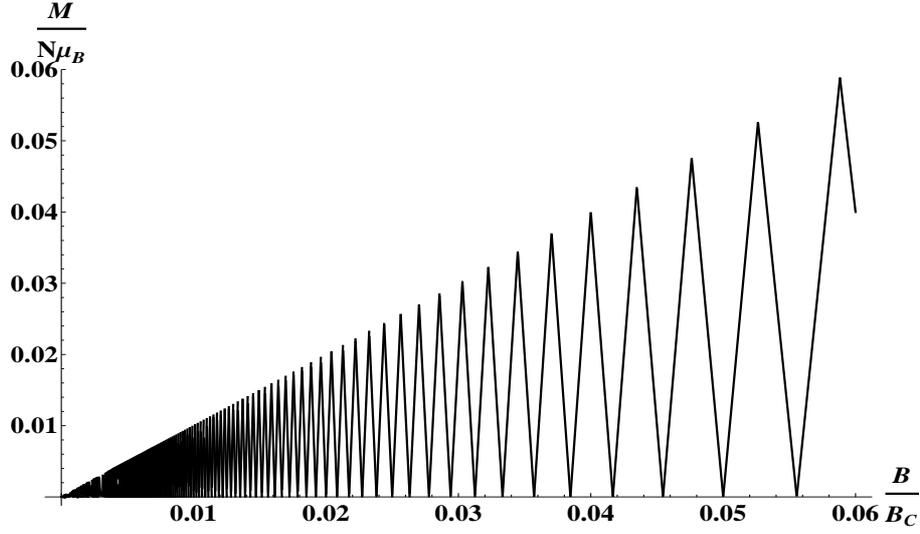}
\caption{Dimensionless magnetization per number of electrons in the region
where spin up and down levels do not mix. }
\label{fig1}
\end{figure}

Let us consider the first case under study: graphene under a magnetic field
with Zeeman effect. In this case, the energy levels can be found and reads%
\begin{equation}
E_{n,s,l}=-s\hbar \omega _{z}+l\hbar \omega _{L}\sqrt{n}  \label{n2}
\end{equation}%
being $s=\pm 1$ for spin up and down respectively, $n=0,1,2,...$ is the
Landau level index and $l=1$ for the conduction band and $l=-1$ for the
valence band. For the conduction band, electrons will start filling the
lowest levels, so in the case in which $\left\vert \hbar \omega
_{z}\right\vert <\frac{1}{2}\left\vert \hbar \omega _{L}\right\vert $, that
is, when the Zeeman splitting is lower than half the separation between
consecutive Landau levels, then the filling will be done considering the
Landau index first and then the spin index.

\begin{figure}[tbh]
\begin{minipage}{0.48\linewidth}
\includegraphics[width=84mm,height=57mm]{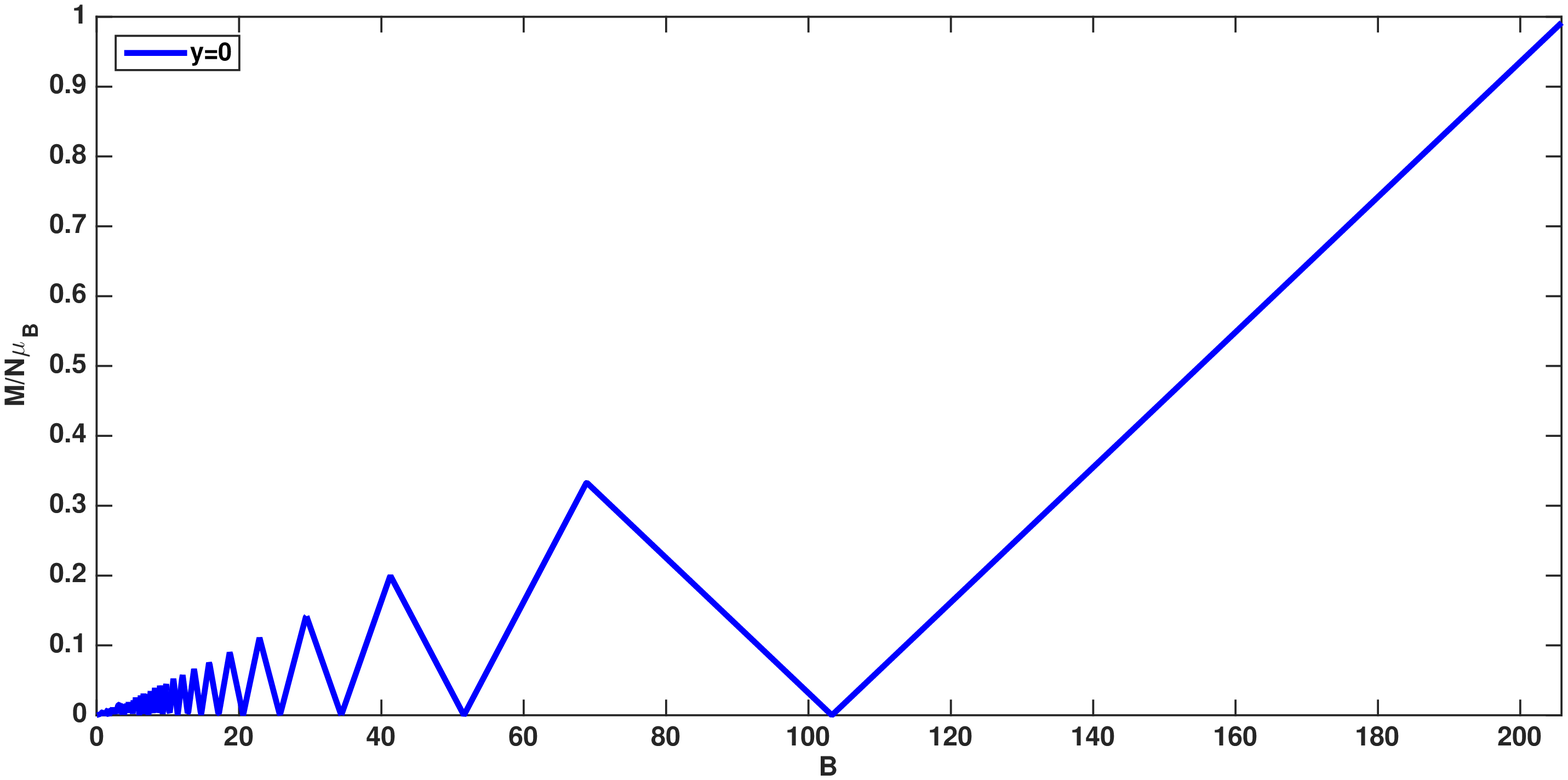} 
\label{proA}
\end{minipage}
\hspace{0.07cm} 
\begin{minipage}{0.5\linewidth}
\centering
\includegraphics[width=86mm,height=57mm]{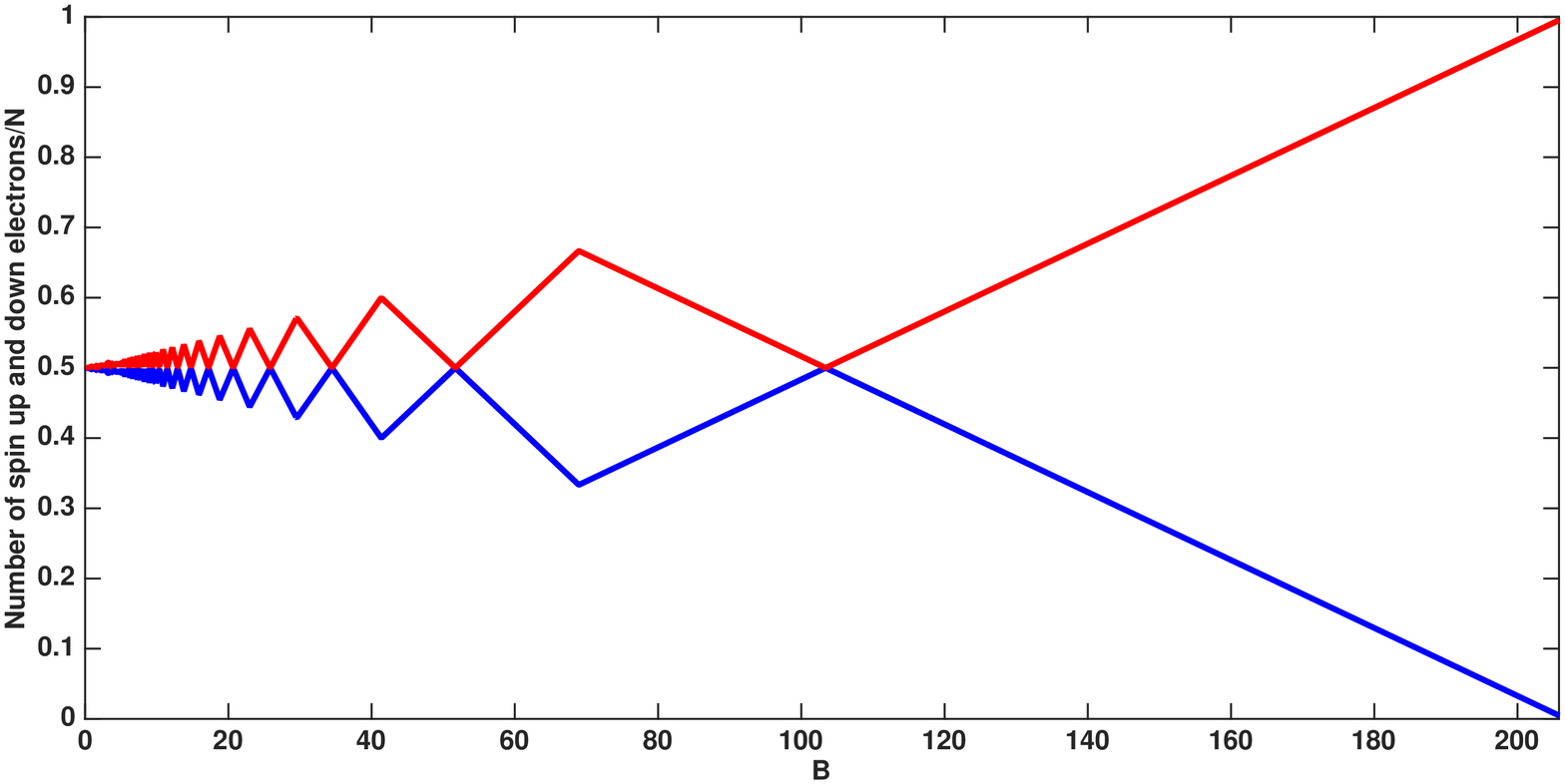}
\label{proB}
\end{minipage}
\caption{Left. Spin magnetization as a function of the magnetic field
strength, where $n_{e}=0.1$nm$^{2}$ and $A=10$nm$^{2}$ without spin-orbit
coupling. Right:\ Number of spin up and down electrons normalized with
respect the number of electrons $N$. }
\label{fig2}
\end{figure}
This implies that if we consider the energy levels in ascending order, there
will be no consecutive Landau levels with same spin filled. This assumption
is fulfilled for normal systems, where $\omega _{L}$ is cyclotron frequency $%
\omega _{L}=\frac{eB}{m}$ and the Landau levels are $\hbar \omega
_{C}(n+1/2) $. 
\begin{figure}[ht]
\label{ fig7} 
\begin{minipage}[b]{0.5\linewidth}
    \includegraphics[width=84mm,height=57mm]{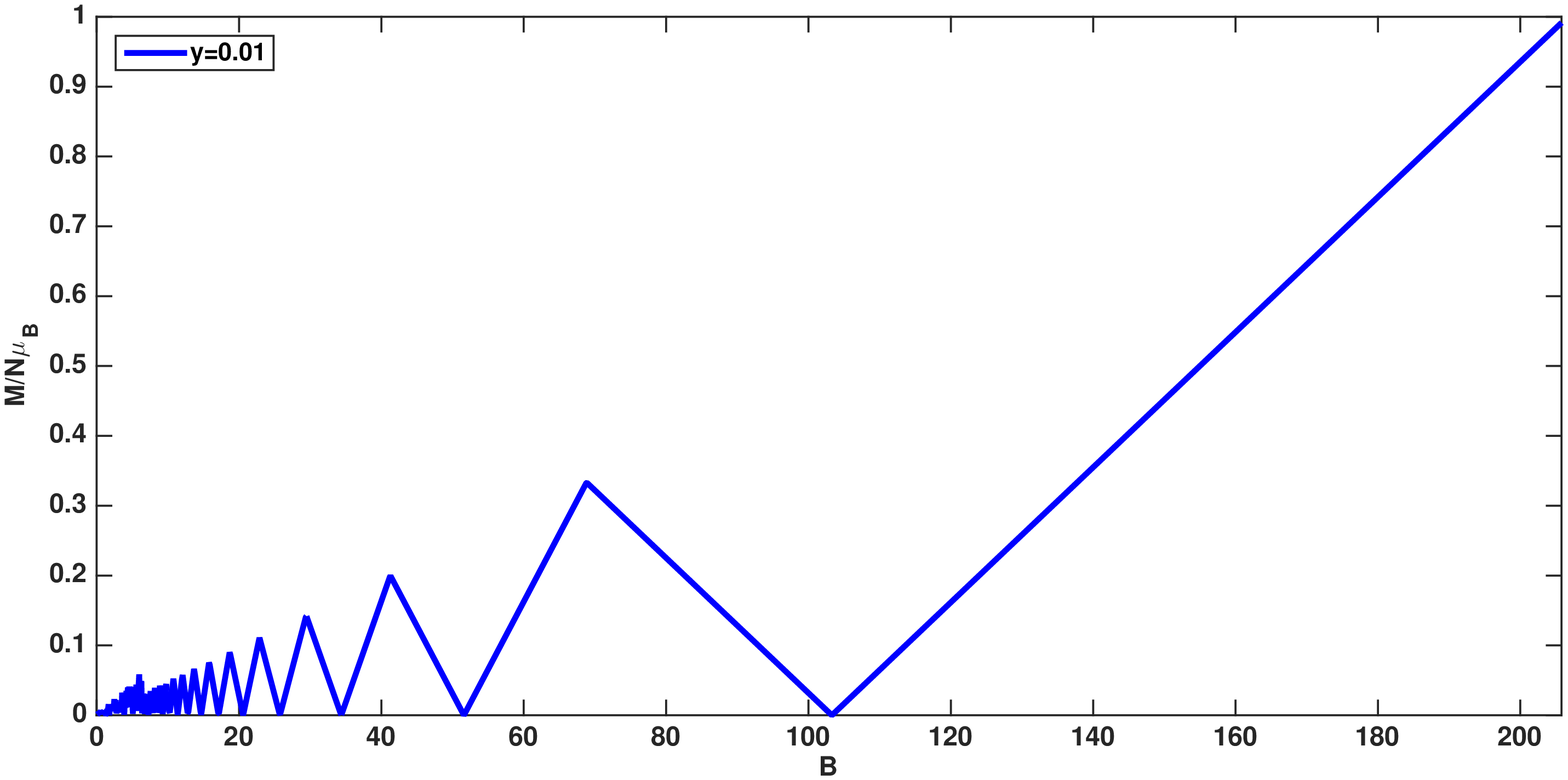} 
   \caption{Spin magnetization for $y=0.01$eV.} 
\label{A1}
  \end{minipage} 
\begin{minipage}[b]{0.5\linewidth}
    \includegraphics[width=84mm,height=57mm]{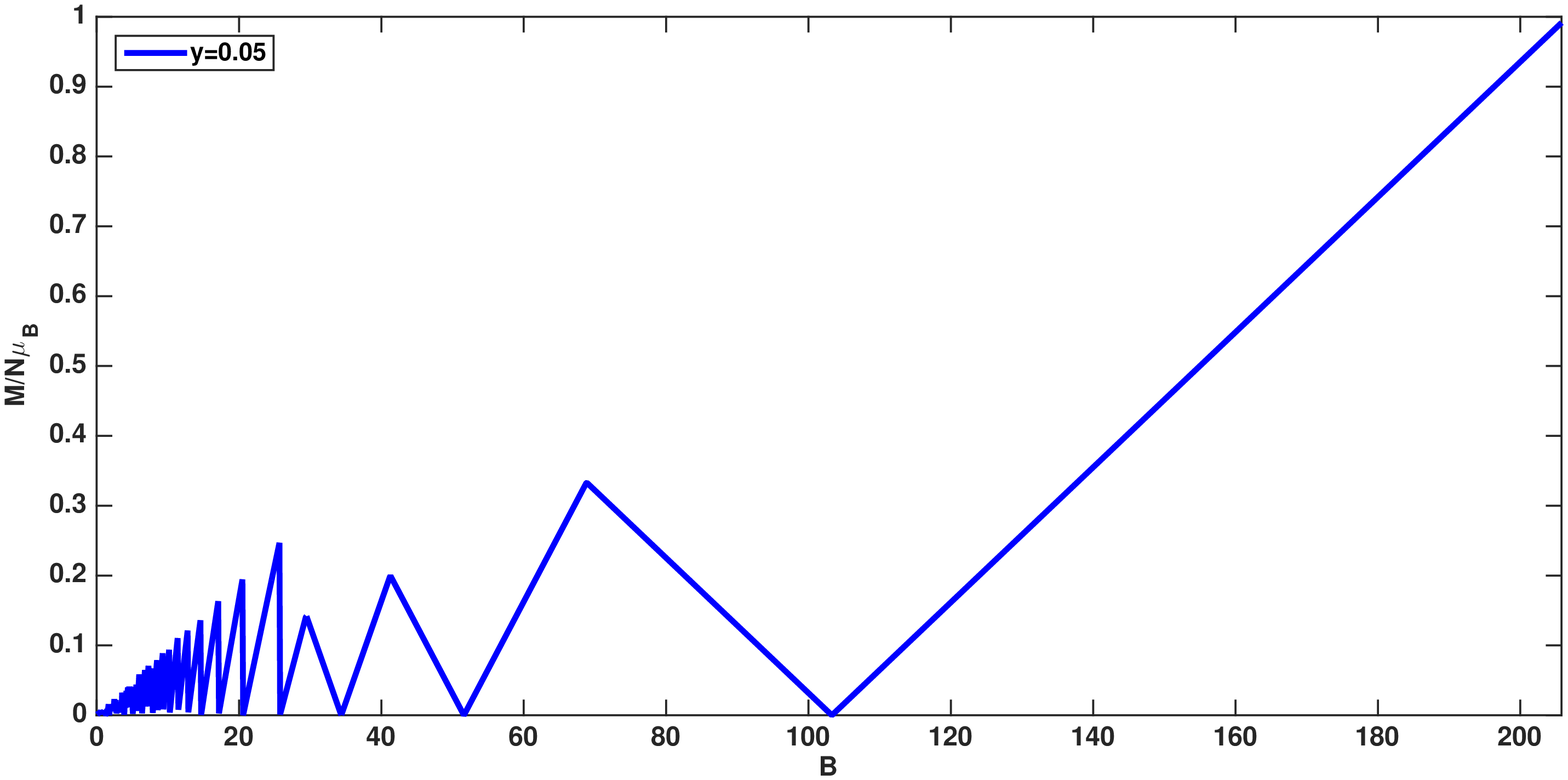}
   \caption{Spin magnetization for $y=0.05$eV.} 
\label{A2} 
  \end{minipage} 
\begin{minipage}[b]{0.5\linewidth}
    \includegraphics[width=84mm,height=57mm]{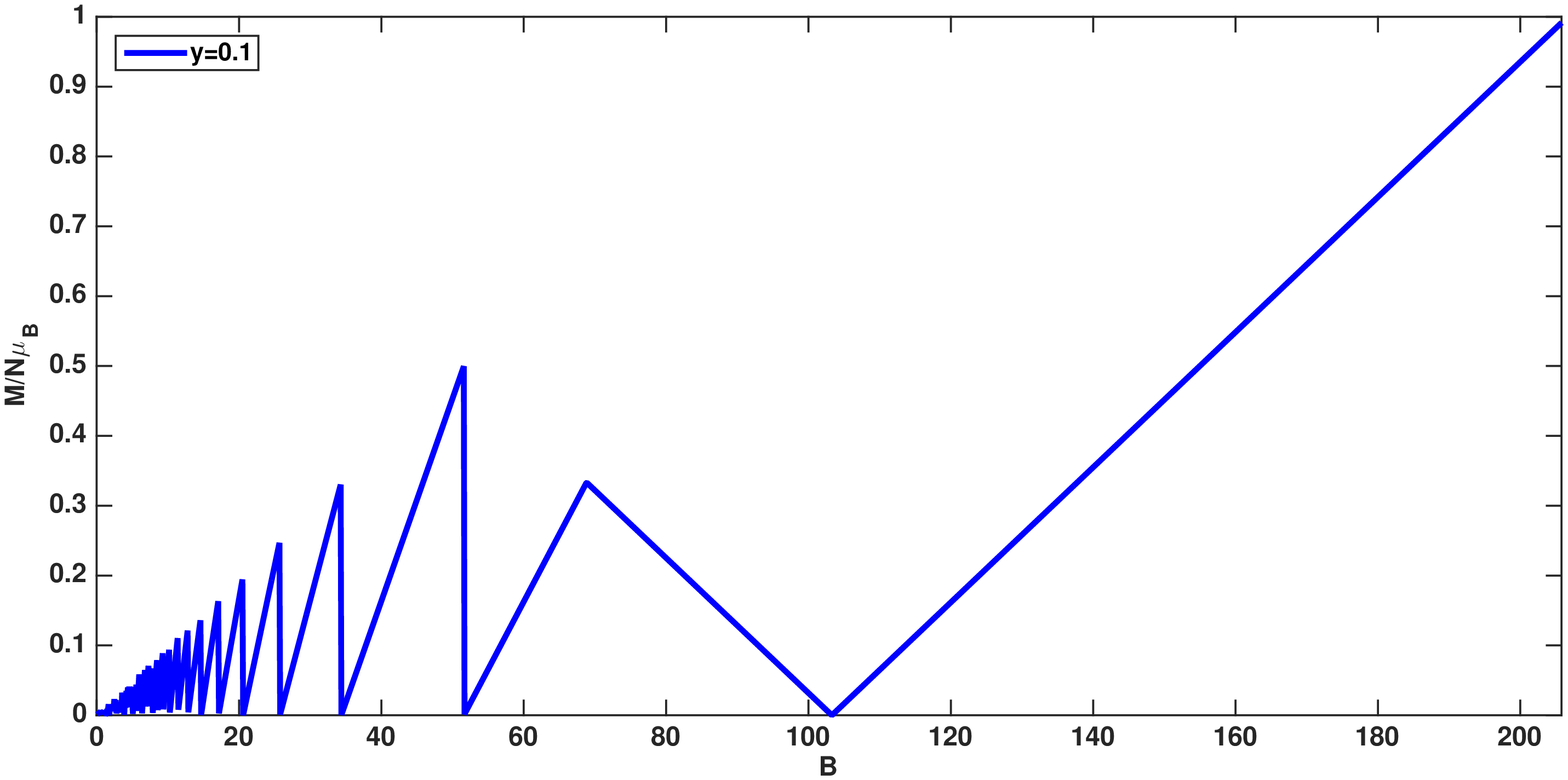} 
   \caption{Spin magnetization for $y=0.1$eV.} 
\label{A3}
  \end{minipage}
\hfill 
\begin{minipage}[b]{0.5\linewidth}
    \includegraphics[width=84mm,height=57mm]{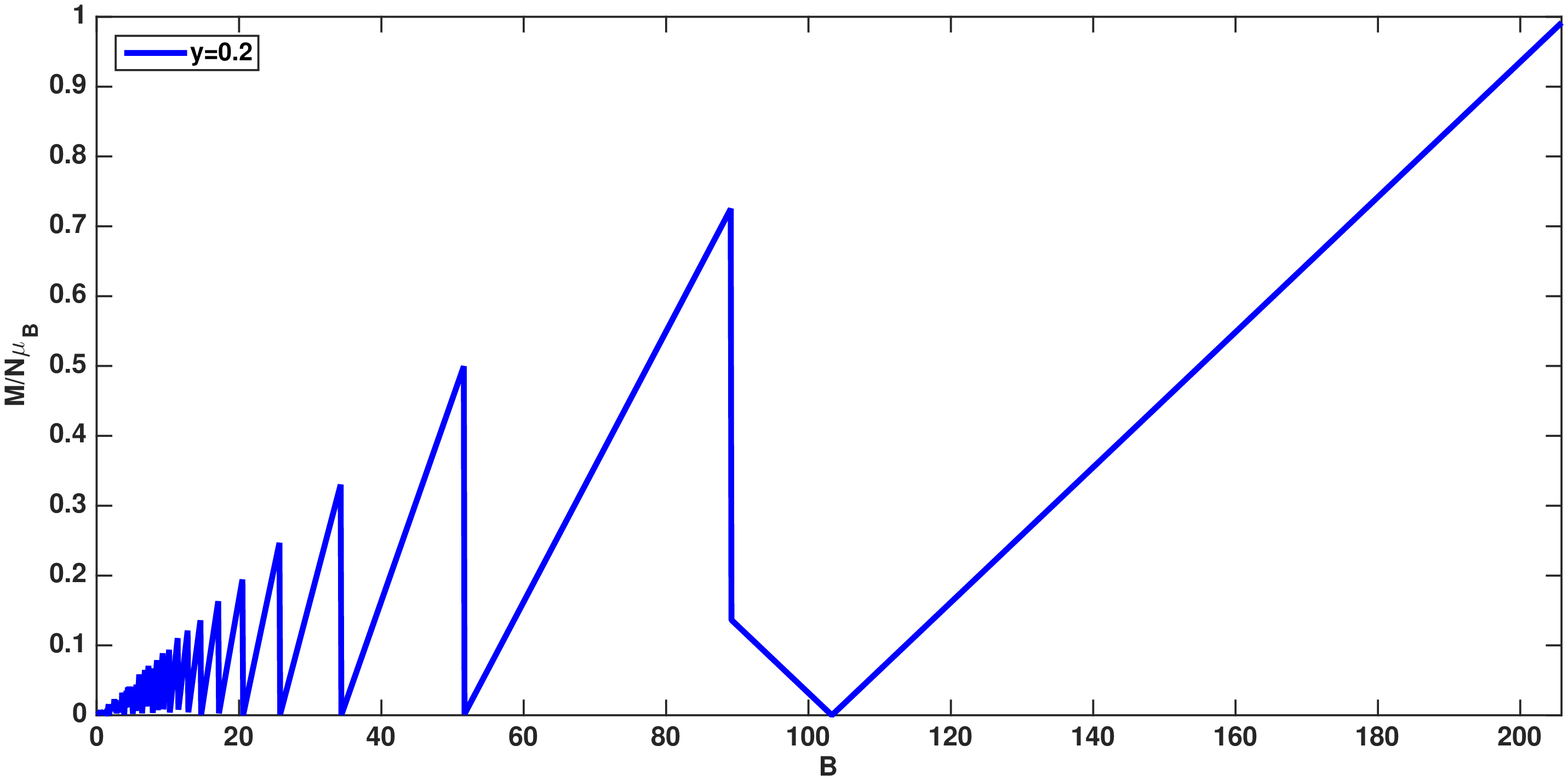} 
   \caption{Spin magnetization for $y=0.2$eV.} 
\label{A4}
  \end{minipage} 
\end{figure}

In this case, because $\omega _{Z}$ is proportional to $B$, then the
condition $\left\vert \hbar \omega _{z}\right\vert <\frac{1}{2}\left\vert
\hbar \omega _{L}\right\vert $ implies that $\frac{g\mu _{B}}{2}<\frac{\hbar
e}{m}$. For graphene, due to the relativistic dispersion relation, $\omega
_{L}$ is proportional to $\sqrt{B}$ and the Landau levels increase as $\sqrt{%
n}$. Then if we write $\hbar \omega _{z}=\gamma _{Z}B$ and $\hbar \omega
_{L}=\gamma _{L}\sqrt{B}$, then the condition for no consecutive spin up or
down filling is given by%
\begin{equation}
B<\left( \frac{\gamma _{L}}{2\gamma _{Z}}\right) ^{2}\left( \sqrt{n+1}-\sqrt{%
n}\right) ^{2}  \label{n3}
\end{equation}

\begin{figure}[tbh]
\begin{minipage}{0.48\linewidth}
\includegraphics[width=84mm,height=57mm]{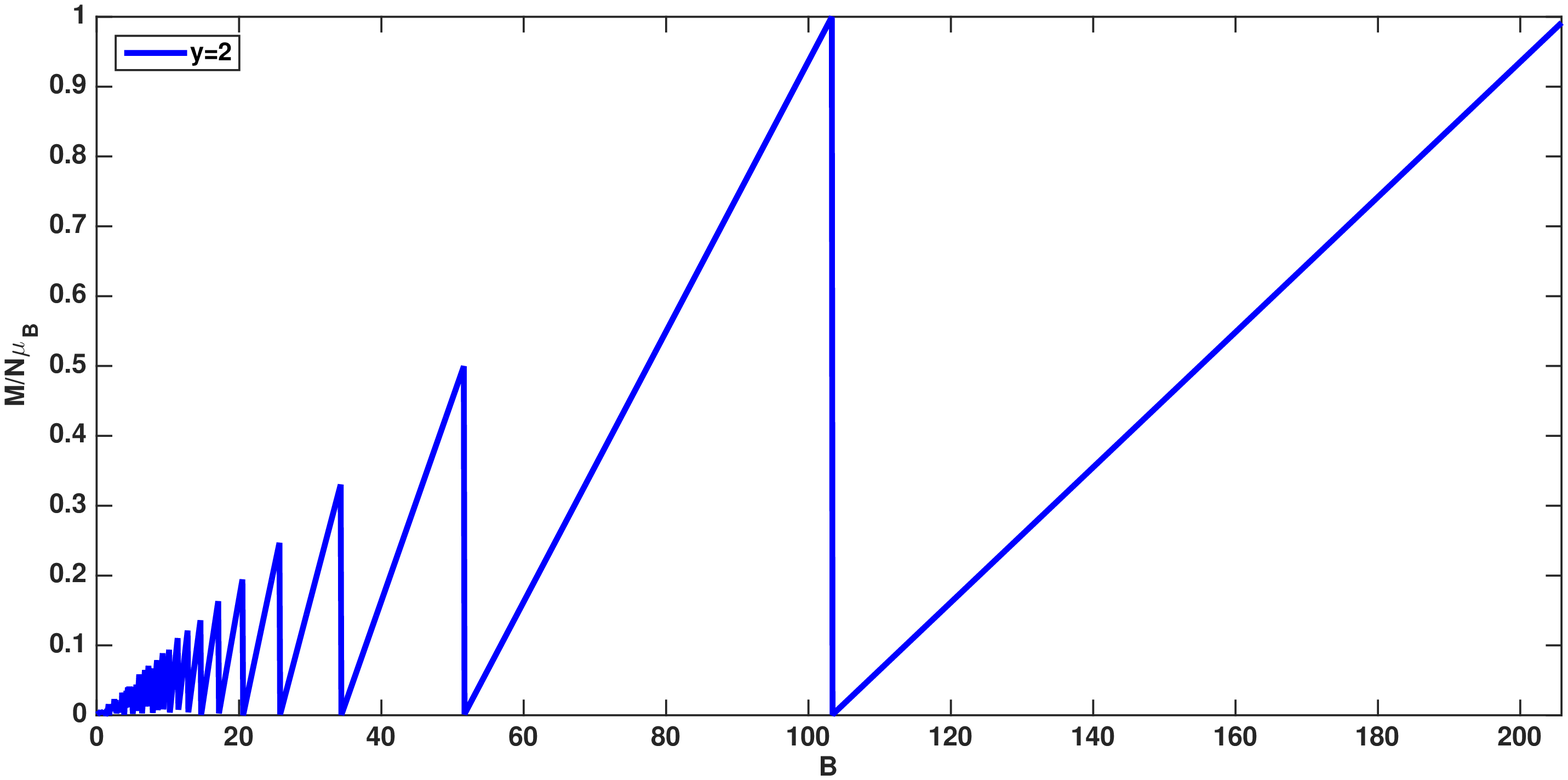} 
\label{proA}
\end{minipage}
\hspace{0.07cm} 
\begin{minipage}{0.5\linewidth}
\centering
\includegraphics[width=86mm,height=57mm]{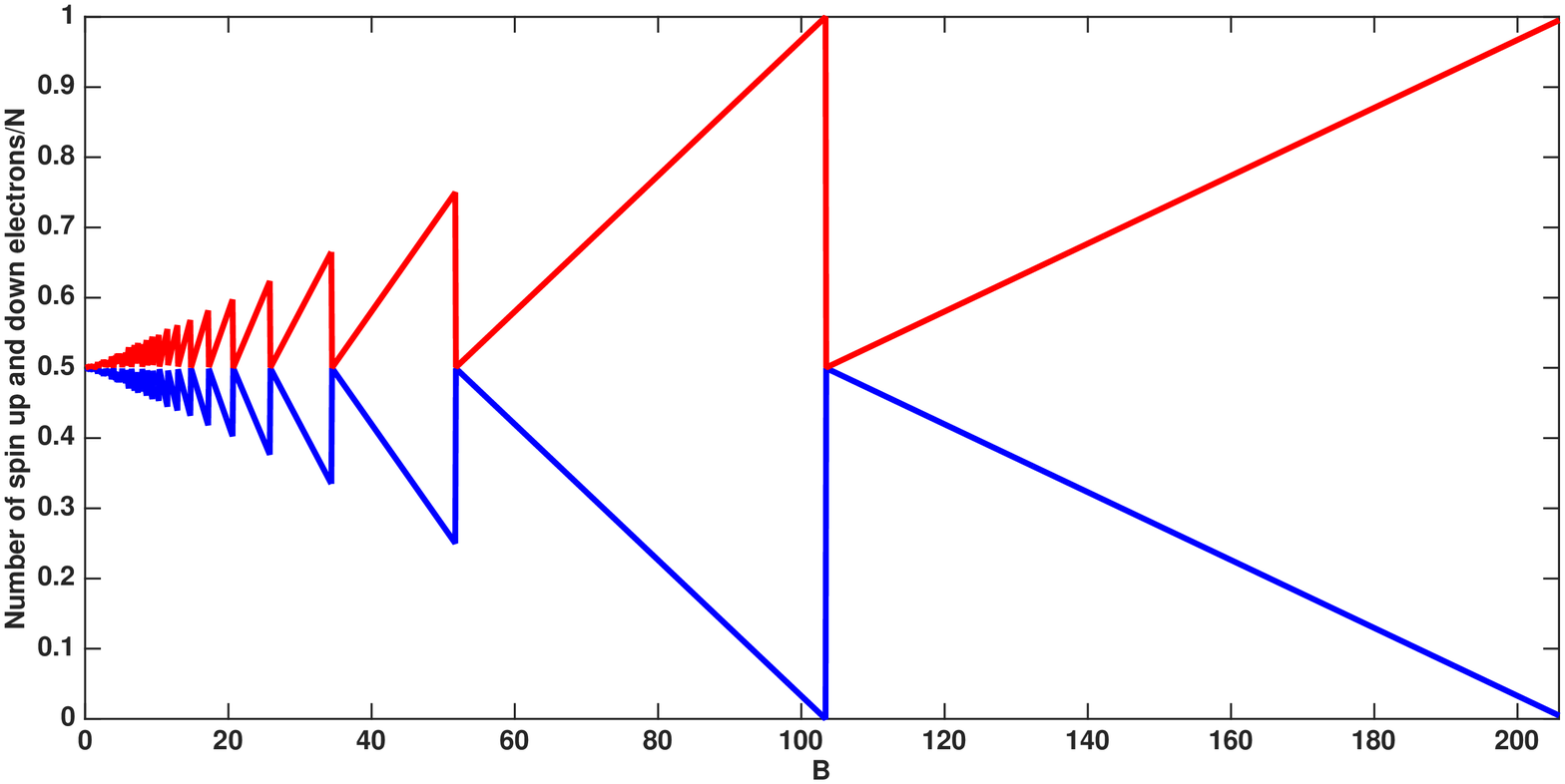}
\label{proB}
\end{minipage}
\label{proB} \label{fig5} \label{figtotal} \label{figtotal1}
\caption{Left. Spin magnetization as a function of the magnetic field
strength, where $n_{e}=0.1$nm$^{2}$ and $A=10$nm$^{2}$. The Rashba
spin-orbit coupling is larger than the critical value where spin
magnetization saturates. Right:\ Number of spin up and down electrons
normalized with respect the number of electrons $N$. }
\label{figtotal11}
\end{figure}
\bigskip For graphene $\frac{\gamma _{L}}{2\gamma _{Z}}\sim \frac{36.29}{%
2\cdot 0.12}\sim 151.2$ (see \cite{sarma2} and \cite{goerbig}).Under this
regime, the ordered energy levels can be written as%
\begin{equation}
E_{p}=-(-1)^{p}\hbar \omega _{Z}+\hbar \omega _{c}\sqrt{\frac{p}{2}-\frac{1}{%
4}(1-(-1)^{p})}  \label{n4}
\end{equation}%
for $p$ even we obtain the spin up levels and for $p$ odd the spin down
levels. The filling factor $v=B_{C}/B$ can be written as $v=q+\theta $,
where $q=\left[ \frac{B_{C}}{B}\right] $, where $\left[ x\right] $ is the
floor function defined as the largest integer less than or equal to $x$ and $%
\theta =\nu -q$. The spin Pauli paramagnetism is given by the remaining term
for the Landau filling which is proportional to $\lambda $%
\begin{equation*}
M=\mu _{B}(N_{\uparrow }-N_{\downarrow })=N\mu _{B}\frac{B}{B_{C}}\left[ 
\frac{1}{2}\left( 1-(-1)^{[\frac{B_{C}}{B}]}\right) +(-1)^{\left[ \frac{B_{C}%
}{B}\right] }\left( \frac{B_{C}}{B}-\left[ \frac{B_{C}}{B}\right] \right) %
\right]
\end{equation*}%
where the difference $(N_{\uparrow }-N_{\downarrow })$ is proportional to $%
\theta $ and the factor $(-1)^{q}$ selects the majority of spin up or down
states in the last Landau level partially filled. In figure \ref{fig1}, the
dimensionless magnetization per electron $M/N\mu _{B}$ is plotted against $x=%
\frac{B}{B_{C}}$.

\begin{figure}[tbp]
\centering\includegraphics[width=115mm,height=70mm]{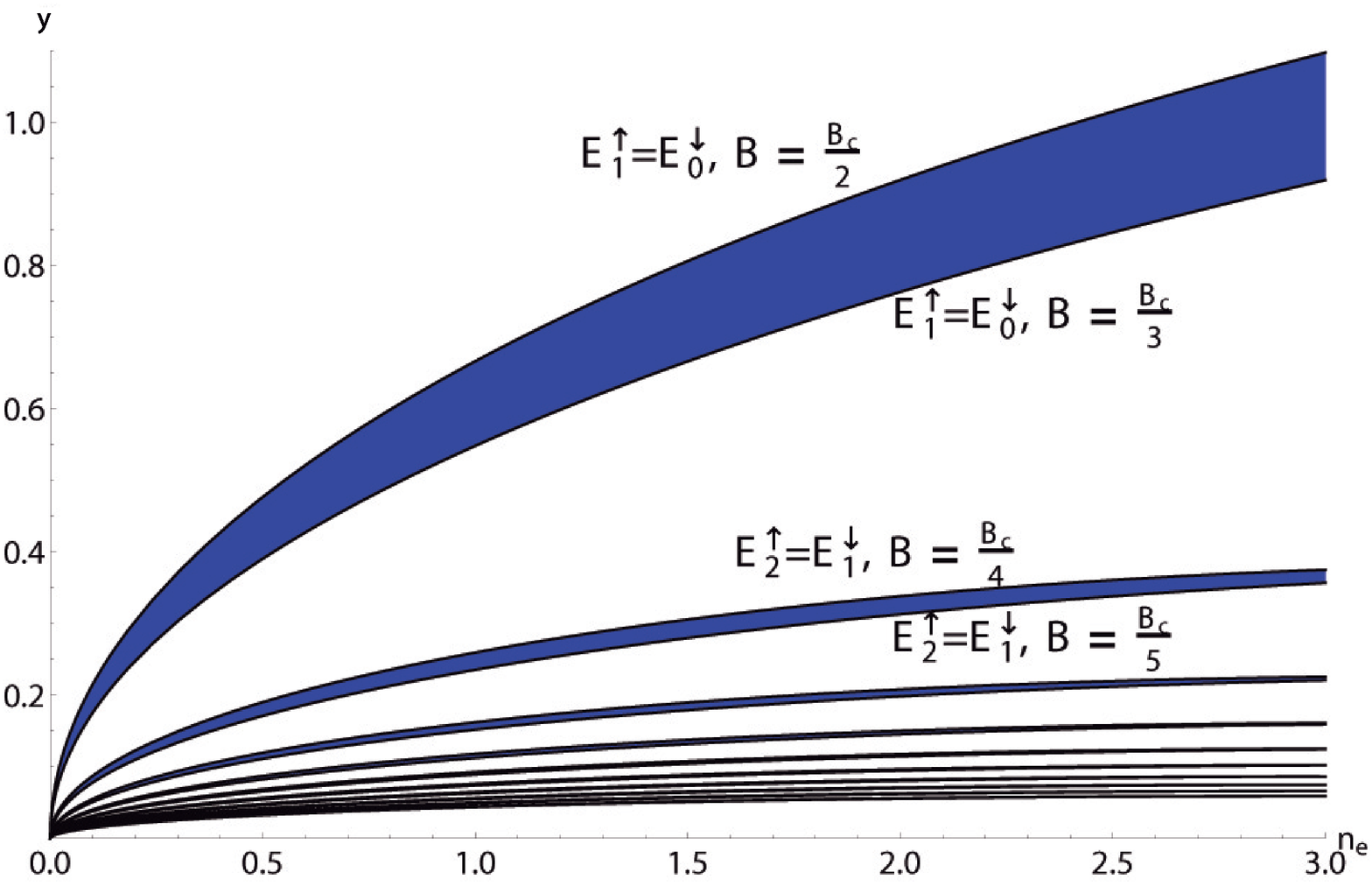}
\caption{Rashba spin-orbit coupling $y$ as a function of the electron
density $n_{e}$. Blue regions between curves indicates those values of $y$
where spin magnetization changes with respect the $y=0$ case. }
\label{filling1}
\end{figure}

\begin{figure}[tbp]
\centering\includegraphics[width=125mm,height=70mm]{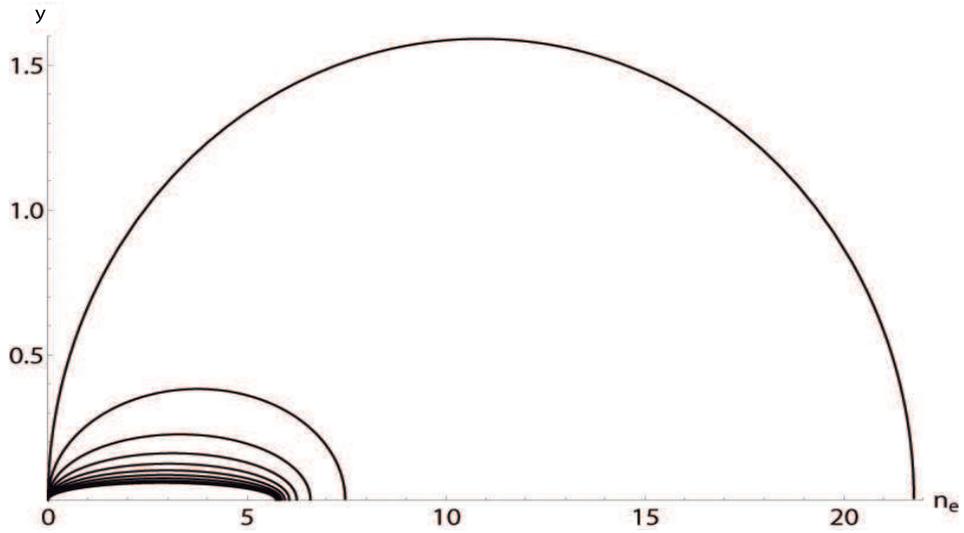}
\caption{Rashba spin-orbit coupling $y$ as a function of the electron
density $n_{e}$. Each curve indicates the values of $y$ where the spin
magnetization jumps from $0$ to $M=\protect\mu _{B}N/2j$ where $j=1,2,3...$. 
}
\label{filling21}
\end{figure}

It must be stressed that condition of eq.(\ref{n3}) depends on $n$ and
because$\sqrt{n+1}-\sqrt{n}<1$ for $n>1$, then for higher Landau levels the
upper bound for $B$ decreases and we must take into account the lowest
Landau level completely filled $n=q=\left[ \frac{B_{C}}{B}\right] $. This
gives the condition for $B$

\begin{equation}
B<\left( \frac{\gamma _{L}}{2\gamma _{Z}}\right) ^{2}\left( \sqrt{\left[ 
\frac{B_{C}}{B}\right] +1}-\sqrt{\left[ \frac{B_{C}}{B}\right] }\right) ^{2}
\label{n4.2}
\end{equation}%
The jumps in the magnetization are well understood in terms of the filling
of the Landau levels up to the Fermi level. Because the Fermi level
indicates the lowest Landau level filled but the degeneration of each level
is not a integer number, then when a level is completely filled, the Fermi
level jumps and in consequence the magnetization. In the second regime,
where $B\geq \left( \frac{\gamma _{L}}{2\gamma _{Z}}\right) ^{2}\left( \sqrt{%
n+1}-\sqrt{n}\right) ^{2}$, the energy levels will show consecutive Landau
levels with the same spin filled. In this case, it is necessary to use
numerical methods to sort the energy levels in ascending order taking into
account if the level belongs to a spin up or down state. In figure \ref{fig2}%
, a sawtooth like behavior for the spin magnetization is obtained which is
given by the spin up and down population, where we have used that $n_{e}=0.1$%
. When $\nu =2$, then $D=N/2$ which implies that the fundamental and first
excited levels are completely filled. For this particular value of $B$, the
first excited level are filled with spin down states. This behavior is shown
in figure \ref{fig2}, where both population are identical for $B=B_{C}/2$.
Between this value and $B=B_{C}$ ($\nu =1$), the spin up population
decreases linearly. The same behavior is expected for $v=l$ where $l=1,2,$%
... For those values of $B$, the spin up and down population are identical.
The spin magnetization peaks follows a linear behavior for high magnetic
fields $M=\frac{B}{n_{e}\phi _{0}}$. For $B>n_{e}\phi _{0}$ the spin
magnetization reaches a plateau given by $M=\mu _{B}N$.

\subsection{Spin-orbit coupling}

When an external electric field is applied perpendicular to the graphene
sheet, the Byckhov-Rashba effect appears (see \cite{kons}). The parameter $y$
that describes this interaction depends linearly with the external electric
field strength. In figures \ref{A1}, \ref{A2}, \ref{A3} and \ref{A4}, a
sequence of plots for different values of $y$ are shown, where $n_{e}=0.1$
and $A=10$nm$^{-2}$ ($N=1$).\footnote{%
A supplementary video has been uploaded where the spin magnetization is
plotted for increasing values of $y$ for the particular case $n_{e}=0.1$nm$%
^{-2}\,$.} As it can be seen in the supplementary material, there are set of
values of $y$ where there are no changes in the magnetization. These set of
values are given when the spin up Landau energy level $n$ is lower than the
spin down energy level $n-1$. For simplicity, let us consider that $\nu =2$.
Then only the two first Landau levels are completely filled when $y=0$. In
this case, because these two levels correspond to the spin up and down $n=0$
Landau levels, then the spin magnetization vanishes because $N_{+}=N_{-}$.
When $y>0$, there is a critical value $y_{c}^{(2)}$ where $E_{1,\uparrow
}<E_{0,\downarrow }$ which implies that the two first Landau levels are spin
up states. Then the magnetization jumps to a constant value $M=\mu _{B}N$.
The condition $E_{1,\uparrow }<E_{0,\downarrow }$ implies that%
\begin{equation}
\text{ \ \ \ \ }\sqrt{2\gamma _{Z}^{2}B^{2}+2y^{2}}\geq \sqrt{2\gamma
_{Z}^{2}B^{2}+2\gamma _{L}^{2}B+y^{2}-\sqrt{4\gamma _{Z}^{2}B^{2}(4B\gamma
_{L}^{2}+y^{2})+y^{4}}}  \label{q1}
\end{equation}%
and the solutions reads%
\begin{equation}
y^{2}\geq \frac{\gamma _{L}^{2}}{2}B-2\gamma _{Z}^{2}B^{2}  \label{q2}
\end{equation}%
which gives two solutions, one for positive values of $y$ which is the case
of interest. In turn, when $\nu =3$, then $D=N/3$, then only the first three
Landau levels are completely filled. For $y=0$, there is no spin mixing and
therefore the magnetization is positive and maximum. When $y>0$ and in
particular when $y>y_{c}^{(3)}$ then $E_{1,\uparrow }<E_{0,\downarrow }$,
but in this case, when we increase the magnetic field strength, the
degeneration increases, which implies that only the two first Landau levels
are filled completely, which correspond to spin up states. Then, the
magnetization increase as $B$ does until we reach $B=\frac{B_{C}}{2}$. These
considerations imply that there is a critical value $%
y_{c}^{(3)}<y<y_{c}^{(2)}$ defined from the equation $E_{1,\uparrow
}=E_{0,\downarrow }$ when $B=\frac{n_{e}\phi _{0}}{2}$ for $y^{(2)}$ and
when $B=\frac{n_{e}\phi _{0}}{3}$ for $y_{c}^{(3)}$ where the magnetization
changes with respect the case $y=0$. The regions for $%
y_{c}^{(j+1)}<y<y_{c}^{(j)}$ and $n_{e}$ where the magnetization changes can
be obtained from the condition $E_{j+1,\uparrow }=E_{j,\downarrow }$ when $B=%
\frac{n_{e}\phi _{0}}{j}$ for $y_{c}^{(j)}$ and $B=\frac{n_{e}\phi _{0}}{j+1}
$ for $y_{c}^{(j+1)}$. In figure \ref{filling1} these set of values are
shown. For small values of $y$ the regions tends to a continuum when $%
j\rightarrow \infty $. In turn, in figures \ref{figtotal11}, the spin
magnetization as a function of $B$ is shown when $y>y_{c}^{(1)}$, where the
magnetization saturates. The set of magnetic fields $B_{j}=\frac{n_{e}\phi
_{0}}{2j}$ are particularly important. For these values and 
\begin{equation}
y_{c}^{(j)}=\sqrt{\frac{\frac{n_{e}\phi _{0}}{2j}(\gamma _{L}^{2}-4\gamma
_{Z}^{2}\frac{n_{e}\phi _{0}}{2j})^{2}-16(j-1)\gamma _{Z}^{2}\gamma _{L}^{2}(%
\frac{n_{e}\phi _{0}}{2j})^{2}}{4(j-1)\gamma _{L}^{2}+2\gamma
_{L}^{2}-8\gamma _{Z}^{2}\frac{n_{e}\phi _{0}}{2j}}}  \label{an1}
\end{equation}%
which is found by computing $E_{j+1,\uparrow }<E_{j,\downarrow }$ and
replacing $B_{j}=\frac{n_{e}\phi _{0}}{2j}$, the magnetization jumps from $%
M=0$ to $M_{j}=\mu _{B}N/2j$ (see figure \ref{filling21}). There are
infinite mixes of Landau spin up and down states that are defined through
the inequality $E_{n+r,\uparrow }\leq E_{r,\downarrow }$ which implies that%
\begin{equation}
y^{2}\geq \frac{16n\gamma _{Z}^{2}\gamma _{L}^{2}-(r\gamma _{L}^{2}-4\gamma
_{Z}^{2})^{2}}{8\gamma _{Z}^{2}-2(2n+r)\gamma _{L}^{2}}  \label{an2}
\end{equation}%
These set of values are located between the regions described in figure \ref%
{filling1}.

These results are of importance to develop two-dimensional supertlattices
structures by using graphene or silicene, which supports charge carriers
behaving as massless Dirac fermions with graphene-like electronic band
structure (\cite{liu1} and \cite{cahan}). Spin-orbit interaction in silicene
can be 1000 times larger that of graphene, which implies that quantum spin
Hall effect in silicene is experimentally accesible. Considering two
semi-infinite graphene ribbons, that are employed as source and drain, and a
drain voltage that introduces charge carriers with a concentration $n_{e}$
it is possible to develop a superlattice of different graphene samples with
specific values of $y$ in series. By fixing the magnetic field strength in
any of the critical values $B_{j}=\frac{n_{e}\phi _{0}}{2j}$, the possible
values of $y$ can be located over the curves in figure \ref{filling21} in
such a way to increase the spin polarized density of states across the
superlattice. Different configurations can be obtained by developing
superlattice with alternating constant electric fields in order to obtain
spin flipping, which can be used as a spin down or up filter \cite{pour}. By
applying Landauer-B\"{u}ttiker formalism, the electronic behavior of Dirac
fermions in the superlattice can be studied. The conductance will shows
resonant tunneling behavior depending on the number of barriers and barrier
width \cite{xiao2}. In turn, the spin polarization lifetime can be
controlled by the applied electric field that controls Rashba spin-orbit
coupling \cite{an2} and \cite{dali}. It has been shown the angular range of
the spin-inversion can be efficiently controlled by the number of barriers 
\cite{sattari}. Magnons can be obtained in the superlattice structure by
combining alternating applied electric fields, where the wavelength can be
accomodated by the width of the middle graphene samples \cite{haja}. It is
source of future works to develop graphene superlattice with different
configurations of Rashba spin-orbit couplings and external magnetic fields.

Finally, to further explore the magnetic activation by changing the RSOC, we
can consider the following setup:\ $N=2$, that is, two conduction electrons
only, $B=\frac{B_{c}}{2}$, where $D=1$ which implies that we have to
consider the first two Landau levels only. When $y=0$, one electron occupy
the ground state with spin up and the second electron the ground state with
spin down which implies that the spin magnetization is zero. Suppose that we
consider the action of a fast sudden perturbation in the system where the
RSOC changes to $y=y_{c}^{(1)}$, where as we said before, $y_{c}^{(1)}$ is
such that $E_{1,\uparrow }=E_{0,\downarrow }$ which is given in eq.(\ref{q2}%
). Then, because the first excited spin up level is below the ground state
with spin down, the first two Landau levels filled are with spin up, which
implies that $M$ jumps to $M=2\mu _{B}$. The transition amplitude is given
by the inner product of both eigenstates. To obtain this value we can
consider the anti-symmetric state for $y=0$%
\begin{equation}
\left\vert \psi _{M=0}\right\rangle =\frac{1}{\sqrt{2}}\left[ \left\vert
\varphi _{0,\uparrow }^{y=0}\right\rangle \otimes \left\vert \varphi
_{0,\downarrow }^{y=0}\right\rangle -\left\vert \varphi _{0,\downarrow
}^{y=0}\right\rangle \otimes \left\vert \varphi _{0,\uparrow
}^{y=0}\right\rangle \right]   \label{d1}
\end{equation}%
and the anti-symmetric state for $y=y_{c}^{(1)}$%
\begin{equation}
\left\vert \psi _{M=2\mu _{B}}\right\rangle =\frac{1}{\sqrt{2}}\left[
\left\vert \varphi _{0,\uparrow }^{y=y_{c}^{(1)}}\right\rangle \otimes
\left\vert \varphi _{1,\uparrow }^{y=y_{c}^{(1)}}\right\rangle -\left\vert
\varphi _{1,\uparrow }^{y=y_{c}^{(1)}}\right\rangle \otimes \left\vert
\varphi _{0,\uparrow }^{y=y_{c}^{(1)}}\right\rangle \right]   \label{d12}
\end{equation}%
where we can note that the state for the second electron is spin up. where
we are considering that both electrons are not interacting. Each vector $%
\left\vert \varphi _{n,s}^{y=0}\right\rangle $ can be computed by
considering the diagonalization of the Hamiltonian of eq.(\ref{a1}). This
eigenfunctions are computed in \cite{jsaeuro} and \cite{jsa2} and we can
write $\left\vert \varphi _{0,\uparrow }^{y=0}\right\rangle =\frac{1}{\sqrt{%
2L}}\left( 
\begin{array}{cccc}
\phi _{0} & \phi _{0} & 0 & 0%
\end{array}%
\right) $, $\left\vert \varphi _{0,\downarrow }^{y=0}\right\rangle =\frac{1}{%
\sqrt{2L}}\left( 
\begin{array}{cccc}
0 & 0 & \phi _{0} & \phi _{0}%
\end{array}%
\right) $, $\left\vert \varphi _{0,\uparrow }^{y=y_{c}^{(1)}}\right\rangle =%
\frac{1}{\sqrt{2L}\sqrt{\left\vert \beta _{1}\right\vert ^{2}+1}}\left( 
\begin{array}{cccc}
\beta _{1}\phi _{0} & 0 & 0 & \phi _{0}%
\end{array}%
\right) $ and $\left\vert \varphi _{1,\uparrow
}^{y=y_{c}^{(1)}}\right\rangle =\frac{1}{\sqrt{2L}\sqrt{\left\vert \alpha
_{1}\right\vert ^{2}+\left\vert \alpha _{2}\right\vert ^{2}+\left\vert
\alpha _{3}\right\vert ^{2}+1}}\left( 
\begin{array}{cccc}
\alpha _{1}\phi _{1} & \alpha _{2}\phi _{0} & \alpha _{3}\phi _{1} & \phi
_{0}%
\end{array}%
\right) $, where $\phi _{n}(\xi )=\frac{\pi ^{-1/4}}{\sqrt{2^{n}n!}}e^{-%
\frac{1}{2}\xi ^{2}}H_{n}(\xi )$ where $H_{n}$ are the Hermite polynomials
of order $n$ and $\xi =\xi =\frac{y}{l_{B}}-l_{B}k$, where $l_{B}=\sqrt{1/eB}
$ is the magnetic length and $k$ is the wavevector in the $x$ direction. $2L$
is the total length of the graphene sample in the $x$ direction. 
\begin{figure}[tbp]
\centering\includegraphics[width=125mm,height=70mm]{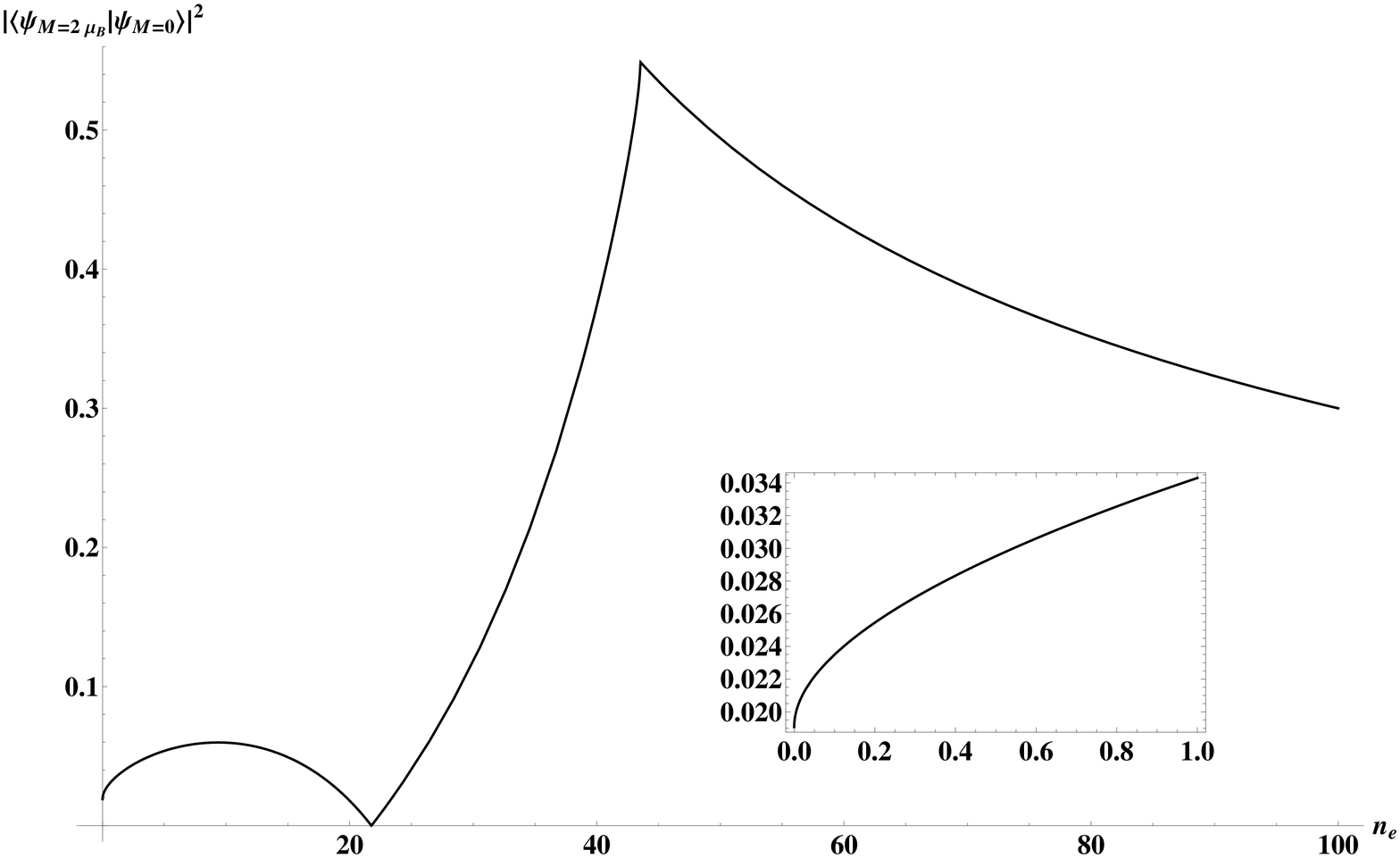}
\caption{Transition amplitude for magnetic activation by a sudden
perturbation $y=0\rightarrow y=y_{c}^{(1)}$ as a function of electron
density $n_{e}$. }
\label{trans}
\end{figure}
The coefficients $\beta _{1}$, $\alpha _{1}$, $\alpha _{2}$ and $\alpha _{3}$
are obtained through the diagonalization of the Hamiltonian and reads%
\begin{equation}
\beta _{1}=\frac{i\left( \sqrt{2}\gamma _{Z}B+\sqrt{\gamma _{L}^{2}B-2\gamma
_{Z}^{2}B^{2}}\right) }{\sqrt{\gamma _{L}^{2}B-4\gamma _{Z}^{2}B^{2}}}
\label{d2}
\end{equation}%
\begin{equation}
\alpha _{1}=\frac{i\left( \gamma _{L}^{2}B+2\sqrt{2}\gamma _{Z}B\sqrt{\gamma
_{L}^{2}B-2\gamma _{Z}^{2}B^{2}}\right) }{\sqrt{\gamma _{L}^{2}B-2\gamma
_{Z}^{2}B^{2}}\left( -\sqrt{2}\gamma _{Z}B+\sqrt{\gamma _{L}^{2}B-2\gamma
_{Z}^{2}B^{2}}\right) }  \label{d3}
\end{equation}%
\begin{equation}
\alpha _{2}=\frac{2i\sqrt{2}\gamma _{L}^{2}B\sqrt{\gamma _{L}^{2}B-4\gamma
_{Z}^{2}B^{2}}}{2\gamma _{L}^{2}B-4\sqrt{2}\gamma _{Z}B\sqrt{\gamma
_{L}^{2}B-2\gamma _{Z}^{2}B^{2}}}  \label{d4}
\end{equation}%
\begin{equation}
\alpha _{3}=\frac{\gamma _{L}^{2}B}{2-\gamma _{Z}B+\sqrt{\gamma
_{L}^{2}B+2\gamma _{Z}^{2}B^{2}}}  \label{d5}
\end{equation}%
where we have to replace $B=\frac{B_{c}}{2}$ and $y=y_{c}^{(1)}$. In figure %
\ref{trans}, the probability amplitude $\left\vert \left\langle \psi
_{M=0}\mid \psi _{M=2\mu _{B}}\right\rangle \right\vert ^{2}$ is plotted as
a function of the electron density $n_{e}$. It is shown that the transition
tends to zero as $n_{e}\rightarrow 0$ and a non-trivial zero for $n_{e}\sim
21$nm$^{-2}$. The curve in this region follows the behavior of the curve $%
y=y_{c}^{(1)}$ as it can be seen in figure \ref{filling21}, although it is
not exactly a semicircle. For higher values of $n_{e}$, the probability
amplitude tends to zero. Then it is possible to increase the spin magnetic
transition for the specific value, where the probablity amplitude is maximum
for $n_{e}$. In this model we have not consider more than two electrons
because in this case we should take into account the interactions between
electrons in the same state, which can be done by considering the Laughlin
wavefunctions \cite{laugh} and the subtleties introduced by the fractional
quantum Hall effect.

\section{Conclusions}

In this work we have examined the spin magnetization in pristine graphene
with spin-orbit coupling and Zeeman splitting. The magnetization has been
found as a function of the electron density, the magnetic field strength and
the Rashba spin-orbit coupling parameter. We have derived and compared the
maximum and minimum of the magnetization with and without spin-orbit
coupling showing that for certain regions of the spin-orbit coupling
parameter $y$ and certain values of the electron density $n_{e}$, the
magnetization jumps from zero to a constant value given $M_{j}=\mu _{B}N/2j$
being $N$ the number of electrons and $j=1,2,3...~$being an integer number
that controls when a spin up state with Landau level $j+1$ is lower than the
spin down state with Landau level $j$. Morever, these regions for $y$
follows a complex relation with $n_{e}$ that numerically shows certain
values of $y$ where magnetization is not altered by $y$. These results show
that it is possible to obtain spin filters or spin oscillations by
considering different superlattice configurations, where the inner graphene
samples can alternate the applied electric field in such a way to fix the
magnetic field strength in those values where the Landau spin up and down
states mix. Finally, we have studied the probability amplitude for a sudden
change of the Rashba spin-orbit coupling from the non-magnetic state $M=0$
to $M=2\mu _{B}$ when there are two electrons, $B=\frac{n_{e}\phi _{0}}{2}$
and $y=y_{c}^{(1)}$, showing that the transition has a a peak for $n_{e}>0$
and a non-trivial zero.

\section{Acknowledgment}

This paper was partially supported by grants of CONICET (Argentina National
Research Council) and Universidad Nacional del Sur (UNS) and by ANPCyT
through PICT 1770, and PIP-CONICET Nos. 114-200901-00272 and
114-200901-00068 research grants, as well as by SGCyT-UNS., J. S. A. and L.
S. are members of CONICET., F. E. is a fellow researcher at this institution.

\section{Author contributions}

All authors contributed equally to all aspects of this work.

\end{document}